\documentstyle[12pt]{article}
\newcommand{\be}{\begin{equation}}
\makeatletter
\@addtoreset{equation}{section}
\makeatother

\newcommand{\ee}{\end{equation}}
\newcommand{\bea}{\begin{eqnarray}}
\newcommand{\eea}{\end{eqnarray}}
\newcommand{\nn}{\nonumber}

\begin{document}

\begin{flushright}
\bf JINR Preprint E2-94-332
\end{flushright}
\begin{center}

{\large \bf {ZERO MODES OF GAUSS' CONSTRAINT IN \\[3mm]
GAUGELESS REDUCTION OF YANG - MILLS \\[3mm]
THEORY}\/ }

\vspace*{0.8cm}
{\sc {A. Khvedelidze,~\footnotemark ~  V. Pervushin} }

{Bogoliubov Theoretical Laboratory, Joint Institute for Nuclear Research,\\
Dubna, Russia}

\end{center}

\footnotetext{Permanent address : Tbilisi Mathematical Institute, Tbilisi,
Georgia\\ Electronic address: khved@theor.jinrc.dubna.su  }

\vspace*{0.8cm}

\abstract{
The physical variables for pure Yang - Mills theory  in
four - dimensional Minkow\-skian space time are constructed
{\it without using a gauge \-- fixing condition} by the
explicit resolving of the non - Abelian  Gauss constraint
and by the Bogoliubov
transformation
that diagonalizes the kinetic term in reduced action (action on constraint
shell).
As a result, the reduced action is expressed in terms of  gauge invariant
field variables including an  additional
global (only time - dependent) one, describing  zero mode dynamics of
the Gauss constraint.
This additional variable reflects the symmetry  group of topologically
nontrivial transformations remaining after the reduction.
( It gives also the characteristic of the Gribov ambiguity from the
point of view of the gauge fixing method.)

The perturbation theory in terms of quasiparticles with the new
stable vacuum, which is defined through the zero mode configuration, is
 proposed. It is shown, that the averaging of
Green's functions  for quasiparticles
over the global variable leads to  the mechanism of color confinement.
}

\newpage
\vspace*{2.5cm}
\section{Introduction}

The identification of physical degrees of freedom of the non - Abelian
gauge theory is a crucial  point for understanding  physical
phenomena hidden in its structure.
The procedure of identification of physical variables and their separation
>from nonphysical ones has been called the reduction procedure. There are two
ways for the realization of the reduction : {\it  gaugeless \/}{and}
{\it gauge  fixing\/}.
In the former, independent physical variables are constructed by the explicit
resolution of constraints. In this case nonphysical variables disappear
and the remaining gauge invariant  variables describe
a usual unconstrained system.
To avoid the difficulties with the resolution of  complete
constraints, one commonly uses the general method of Dirac ~\cite{Dirac59},
gauge - fixing approach, based on the introduction into the theory of
some new `` gauge constraints '' and on replacement of the Poisson bracket by
Dirac's one.
However, there are also some problems in the Dirac  approach. In particular,
the  gauge - fixing scheme is based on  ``gauge equivalence '' theorem.
The rigorous proof of gauge independence is known
in the assumption of existence of asymptotically free
states of elementary particles~\cite{Slav}, ~\cite{FadSlav}.
The extension of this result to more general cases, including
a nonperturbative one, is quite problematic. In this case to obtain a gauge
invariant
results a greater accuracy it is necessary to take into account nontrivial
boundary conditions for gauge fields \cite{Tav}.
Another problem of gauge - fixing procedure for the non - Abelian gauge theory
is Gribov's ambiguity ~\cite{Gribov} :
there are many equivalent gauge field configurations obeying the same gauge
condition.
After the study of the space of orbits ( the space of  gauge fields modulo
the group of gauge transformations ) ~\cite{Singer} it has been
clear that the gauge - fixing procedure is not quite painless and it needs
more rigorous treatment.
For these reasons from time to time  attempts have been undertaken to
deal with the gaugeless method ~\cite{Pervush1} - \cite{Khved}.

In this article we would like to demonstrate that the gaugeless method
of quantization allows us to describe the gauge invariant content of
Gribov's ambiguity
taking into account the collective variable inherent in gauge theory with
nontrivial topological properties of gauge group.

According to the gaugeless approach ~\cite{Pervush1}, \cite{PervushRiv},
\cite{Khved},  the physical
variables in the Yang - Mills theory is constructed in the following way :
the  non - Abelian  Gauss' constraint is solved with
respect to the nondynamical time component of gauge field, and then
a new gauge invariant variables are constructed
with the help of the Bogoliubov transformation \cite{Bogol} in terms of the
constraint solution.
The solution of Gauss' equation for nondynamical
variable  is definite within the arbitrary time - dependent functions.
These functions for simple cases (e.g., electrodynamics in infinite space time)
are excluded from consideration owing to  boundary conditions for gauge
fields. In the general case these functions arise in the kinetic term of the
constrained action and thus these zero modes are  physical variables.
In particular,  Bogoliubov's  transformation is defined within the arbitrary
time - dependent phase.
The appearance of this phase in the framework of gaugeless method
corresponds to the Gribov ambiguity in the gauge - fixing approach,
where it can be treated as a  collective variable for  the
group of transformations remaining after imposing the gauge condition.

The paper will be organized as  follows.
In Section 2  we shall briefly describe the gaugeless
method by example of QED in the four - dimensional Minkowskian space - time.
Section  3  is  devoted  to  the  construction of physical variables in
(1+3)  dimensional $SU(2)$ Yang - Mills  theory with  zero mode sector of
Gauss' law. The connection is shown of the phase zero mode variable with
the  winding number functional.
 We  prove  a  no - go  theorem  about the local
realization  of  the  representation  of  a homotopy group without the
collective mode,  and show  that the  presence of a zero  mode
leads  to  another   realization
different from the `` instanton '' one  ~\cite{BPST,Fad}.
In the last section
the perturbation theory in terms of quasiparticles around the stable
vacuum, corresponding to zero mode configuration, is proposed.
All  observables are determined after the averaging over
the collective variable.
It is noted, that the degeneration of physical states with respect to
this collective variable can be cause of the color confinement.

\section{Reduction of QED without gauge fixing}

To introduce the  method of gaugeless reduction,  let us begin with
an example of electrodynamics  in the instant form
\begin{equation}
W[A, \psi, {\bar \psi}]\; =\;
\int d^4x \left( \frac{1}{2}[(\partial_0 A^i-\partial^i A_0)^2 - B_i^2]
+ \bar \psi( i{\hat{\partial}} - m) \psi +
e J_\mu A^\mu \right),
\end{equation}
\[
B_i=\epsilon_{ijk}\partial^j A^k\; .
\]
In the  action (2.1)  in accordance with the choice of time axis
\(\eta \cdot x = x_0 \),
the time component of vector field is distinguished.
The main point of gaugeless reduction for electrodynamics is
to resolve  explicitly the  Lagrangian constraint
\begin{equation}
\Delta A_0\;=\;\partial^i\partial_0 A_i + J_0.
\end{equation}
Within the zero modes of operator \(\Delta \) (explanation will be given
in the next section ) we can write down the solution for (2.2) in the form
of the following decomposition of \(A_0\):
\be
A_0\;=\; A^{tr}_0\;+\;A^{J}_0\;, \quad A^{tr}_0\;=
\frac{1}{\Delta} \left(\partial_0 \partial_j A^j\right)\;, \quad
\;A^{J}_0\;=\;\frac{1}{\Delta} J_0,
\ee
where \(A^{tr}\) is varied under the gauge transformations
\[ A_\mu \;\to\; {A'}_\mu\;=\;A_\mu\;+\;\partial_\mu\Lambda ,\]
while \( A^{J}_0\) remains invariant.
The kinetic term in eq.(2.1) according to this decomposition can be
diagonalized
with  the help of the Bogoliubov transformation to
the new variables ~\cite{Pervush1}:
\begin{eqnarray}
A_{k}^{I}[A]& =& v[A] ( A_{k} - i \frac{1}{e} \partial_{k} )
( v[A] )^{-1},\nn\\
 \psi^{I}[A]& =& v[A] \psi \,,\,\,
\end{eqnarray}
where
\begin{eqnarray}
v[A] =  \exp \bigl\{ i\int_{}^{t} dt' A_0^{tr} \bigl\}\;=\;
\exp \bigl\{ i\, e \frac{1}{{\Delta}}
\partial^{j} A_{j} \bigr\} \,\,\, .
\end{eqnarray}
These variables are the gauge invariant functionals from initial
gauge fields
\[A^I[A + \partial\Lambda]\;=\;A^I[A]\]
and satisfy, by the construction, the identity
\be
\partial_i A_i^I[A] = 0.
\end{equation}
Note that this identity is the consequence of the explicit resolution
(2.3).  Thus, the functional $A^I[A]$ contains only two observable
transverse fields
\[
A_i^I =\sum_{a=1,2} e_i^a A_a^I
\]
without selecting the Coulomb gauge as the initial supposition.
The initial action (2.1) on the constraint shell (2.2) in terms of the new
variables gets the form:
\begin{equation}
W^{Red}[A^I,\psi^I]=\int d^4x \left[\frac{1}{2}\sum_{a=1,2}(\partial_\mu A_a^I
\partial^\mu A_a^I) + \frac{1}{2}j_{0}^{I} \frac{1}{\Delta}j_{0}^{I} -
j_{i}^{I} A_{i}^{I} +
{\bar {\psi}}^{I} (i{\hat \partial} - m){\psi}^{I}\right]\;.
\ee
So, we are ready to pass to the Hamiltonian form for our theory
by using the conventional Legandre transformation for physical coordinates
$A_a^I$.  The quantization is achieved by
imposing the  canonical equal-time commutation relation between
conjugate variables:
\[
[A_a^I(x), E^I_b(y)]\; = \;\delta_{ab} \delta^3(x -y) .
\]
One can  write down  the generating functional for Green's function of the
obtained unconstrained system in the form
\be
Z_{\eta}^{Red}[ s^I, {\bar {s}}^I, J^I ]\;=\;\int \prod_{a} DA^I_a D\psi^I
D{\bar \psi}^I
 e^{iW^{Red}[A^I,\psi^I, {\bar \psi}^I] + i S^I },
\ee
with the external source term
\be
S^I\;=\;\int d^4x \left({\bar s}^I
\psi^I + {\bar \psi}^I s^I +J^I_a A^I_a]\right).
\ee

As to about on gauge and relativistic covariance in gaugeless scheme,
there is  subtle realization of Poincare
symmetry \--- mixing of this rigid symmetry with gauge one
\bea
U_L^{-1}\psi (x)  U_L\;=\; S(L)\exp (ie\Lambda(x,L))\psi (Lx), \nn\\
U_L^{-1}A_\mu (x)  U_L\;=\; (L)^\nu_\mu
\exp (ie\Lambda(x,L)) \left( A_\nu (Lx) + \frac{i}{e}\partial_\mu \right)
\exp (-ie\Lambda(x,L)).
\eea
In particular, it has been proved that the infinitesimal Lorentz
transformation of coordinates with  parameters $\varepsilon^k$
corresponds  to the transformation law for physical variables  ~\cite{Han}:
\[
A^I_i[A+\delta_L A]-A_i^I[A]=\delta_L A^I+\partial_i
\Lambda[A^I]\;,
\]
with the conventional Lorentz variation $\delta_L A = LA$
supplemented by the gauge one:
\[
\Lambda[A^I, J]\;=\;\varepsilon^k \frac{1}{\Delta}
\left[ (\partial_0 A_k^I) + \partial_k J_0\right]\;.
\]
This form was interpreted by Heisenberg and Pauli~\cite{HP} (with
reference to the unpublished note by von Neumann) as the transition from the
Coulomb gauge with respect to the time axis in the rest frame
\(\eta_{\mu}^0=(1,0,0,0)\) to the Coulomb gauge with respect to the
time axis in the moving frame
\[
\eta_{\mu}=\eta_{\mu}^0 + \delta_L {\eta_{\mu}}^0 \;=\; {(L\eta^0)}_{\mu}.\]
The Lorentz cova\-ri\-ance of the re\-duced theory was
proved in the quantum theory by B.Zumino \- \cite{Zumino} and means
\be
Z_{L\eta}^{Red}[ s^I, {\bar {s}}^I,J^I ]\;=\;
Z_{\eta}^{Red}[ Ls^I, L{\bar {s}}^I,LJ^I ].
\ee

\subsection{Gauge equivalence theorem}

The usual form of the gauge -- fixing
Faddeev-Popov integral for generating functional of Green functions
in the gauge \( F(A)\; =\; 0\) is
\begin{eqnarray}
Z^{F}[s^F, {\bar s}^F, J^F]\;&=&\;\int \prod_{\mu}DA^F_\mu D \psi^F D{\bar
\psi}^F
\Delta _{FP}^F
\delta (F(A^F)) e^{iW[ A^F,\psi^F {\bar \psi}^F ]\, + \, i S^F},\nn \\
S^F\;&=&\;\int d^4x \left( {\bar s}^F
\psi^F + {\bar \psi}^F s^F +J^F_\mu A^F_\mu]\right)
\end{eqnarray}
where $W$ is the initial action and  $\Delta_{FP}$ is
the Faddeev -- Popov (FP) determinant.

As was proved ~\cite{Slav}, \cite{FadSlav} by the changing  variables of
integration of the type of (2.4), this representation coincides with the
gauge invariant reduced result  (2.8)  only for the choice of the following
form of the source term
\be
s^F =
( v[A^F] )^{-1}s^I\;,\quad
{\bar s}^F = {\bar s}^I( v[A^F] ) \;,\quad J_0^F =0\,,
\;\;\;\;\partial_i J_i^{F} = 0,
\ee
with the gauge transformation \( v[A^F]  \) defined in (2.5).

In the Feynman diagrams for Green function this gauge factor
\( v[A^F]  \) leads to so-called spurious diagrams (SD).
Just these spurious diagrams restore the Feynman rules (FR) for the
reduced functional (2.8).  Thus, one can present the identity
\be
(FR)^F + (SD) \; \equiv \; (FR)^I\;\quad \mbox{ (for Green's functions) }
\ee
as a consequence of independence of the functional integral (2.12) of
the choice of variables.
It can be verified that in  calculation of the elements of $S$- matrix
transition between the elementary asymptotical states these spurious diagrams
disappear  on the  mass shell surface. As a result, we get
\be
(FR)^F \; = \; (FR)^I  \;\;\quad \mbox{(for elementary particles S- matrix ).}
\ee
This statement  is  known as the  gauge equivalence or independence theorem
{}~\cite{Slav}, \cite{FadSlav}.
When the  asymptotic states contain a composite
particle or some collective excitations, the equation (2.15)
is quite problematic and we are sure  only in  the identity (2.14)
\be
(FR)^F + (SD) \; \equiv \; (FR)^I\;\quad \mbox{(For S - matrix with
composite particles).}
\ee

The violation of the  gauge equivalence  theorem (2.15) in this case does
not mean the gauge noninvariance and relativistic noncovariance.
This violation
reflects the nonequivalence of the different definitions of the sources
(2.9) and (2.13) because of nontrivial boundary conditions and residual
interactions forming asymptotical composite, or collective states.In this
context, the notion "gauge", in fact, is the gauge of sources in the
FP - functional integral, but not only the choice of definite Feynman rules.
As the gaugeless scheme  takes into account explicitly the whole physical
information from  constraints, it is more correct to use this \\
gauge - invariant  and relativistic - covariant scheme
for description of composite particles and collective excitations, rather
than the Dirac approach  with an arbitrary relativistic invariant "gauge of
sources".
Below, we would like to demonstrate the preference of the gaugeless scheme
with an example of collective excitations in the Yang - Mills theory.

\section{GAUGELESS REDUCTION of YANG \--- MILLS THEORY}

\subsection{Zero mode of Gauss' law}

Now we pass to the reduction of the Yang - Mills theory with the local $%
SU(2)$ group in
four - dimensional Minkowskian space - time
\begin{equation}
W [\;A_\mu\;]\; =\;- \frac{1}{4} \int d^4x {F^{a}_{\mu\nu}{F_{a}^{\mu\nu}}}%
\;= \frac{1}{2} \int d^4x \left( {F^{a}_{0i}}^2 - {B^{a}_{i}}^2 \right),
\end{equation}
with the usual definitions of non-Abelian  electric tension $F^a_{0i} $
$$
F_{0i}\;=\;\partial_0 A_i^a - {\nabla (A)} ^{ab}_i A_0^b, \;\;\;\; \nabla
^{ab}_i \;=\; \left(\delta^{ab}\partial_i\ + e \epsilon^{acb}
\,A^c_i\right),
$$
and magnetic one $B^a_i$
$$
B^a_i = \epsilon_{ijk}\; \left( \partial_j A^a_k + \frac{e}{2}\,
\epsilon^{abc}\,A^b_j\,A^c_k \right) .
$$
The reduction consists in the explicit resolution of non-Abelian Gauss' law
\begin{equation}
\frac{\delta W}{\delta A^a_0}= 0 \quad \Longrightarrow \quad
\left[\nabla^2\, (A)\right]^{ac} A^c_0 = \nabla^{ac}\,_i\, (A)
\partial_0 A^c_i         \label{eq:Gauss}
\end{equation}
and next in dealing with the initial action (3.1) on the surface of these
solutions
\begin{eqnarray}
W^{Red}\;=\; W [\;A_\mu\;]\;\Bigl\vert_{
\frac{\delta W}{\delta A^a_0}\;=\;0.}
\end{eqnarray}

Let us choose some particular solution of the constraint (\ref{eq:Gauss}) with
the
property
\be
\lim_{\mid {\vec x} \mid~ \to~ \infty } \;a^c_0 (\vec x,t)\;\; = 0\;\;\;\;
\ee
and write down the general solution as a sum of this particular solution
\(a_0 \) and
the general solution \( \Phi \) ( {\it zero mode field})
of the homogeneous equation:
\begin{equation}
A_0^c\;=\; - \Phi^c\:+\; a^c_0\,, \quad \; \quad
 \left[\nabla^2\, (A)\right]^{ac} \Phi^c = 0\,. \label{eq:zero}
\end{equation}
In the next step the QED gauge invariant variables
(2.4) can be generalized as in ref.~\cite{Pervush1}:
\begin{eqnarray}
\hat A_{k}^{I}& =& v^{I} [A] ( \hat A_{k} + \partial_{k} )
( v^{I} [A] )^{-1},\nn\\
\hat \Phi^{I}& =& v^{I} [A] \hat \Phi ( v^{I} [A] )^{-1}\,,
\quad \quad {\hat A}\;=\;e\frac{A^a \tau^a}{2i}\,,
\label{eq:small}
\end{eqnarray}
where the non - Abelian Bogoliubov transformation \( v^{I} [A]\) is given
with the help of the ``good''  solution \( a_0(\vec x,t) \)
\begin{eqnarray}
v^{I} [A] = T \exp \bigl\{ \int_{}^{t} dt'{\hat a}_0\bigl\}\,.
\end{eqnarray}
In terms of these variables the reduced action  takes the form
\begin{eqnarray}
W^{Red} \big[ A^I \Phi^I \big]  = \frac {1}{2}
\int d^4x \left[ \; (\partial_0 A^I_i +
\nabla_i(A^I)\Phi^I)^2 - B_i^2 \right],
\label{eq:red1}
\end{eqnarray}
where the fields
\(\partial_0 A_i^I,\; \Phi^I,\;  B_i\) satisfy the geometrical constraints
\bea
{\nabla}^{ab}_i(A^I)\;\partial_0 A_i^{bI}\;&=&\;0\,, \nn\\
{\nabla}^{ab}_i\,(A^I){\nabla}^{bc}_i\, (A^I)\; {\Phi}^{Ic} \;& = &\; 0\,,
\nn\\
{\nabla}^{ab}_i(A^I)B_i^b(A^I)\;&=&\;0\,.
\eea
It is clear  that  due to (3.9) the reduced action depends on the zero
mode field \(\Phi \) only through the surface terms on spatial infinity
$I_E$ and  $I_\Phi$:
\begin{eqnarray}
I_E &=&\int d^3 x \partial_0 A_i^I\nabla_i \Phi
\equiv \int d^3 x \partial_i (\partial_0 A_i \Phi)\;=\;
\oint ds_i (\partial_0A_i^I \Phi)
\Bigl\vert_{\mid {\vec  x} \mid~~~ \to~~~ \infty }\,,\\
I_\Phi & = \frac{1}{2}&\int d^3 x (\nabla_i \Phi)^a
(\nabla_i \Phi)^a \equiv \frac{1}{4} \int d^3 x \Delta (\Phi^a)^2
\;=\;
\frac{1}{4}\oint ds_i \partial_i (\Phi)^2
\Bigl\vert_{\mid {\vec  x} \mid~~~ \to~~~ \infty }\,,
\end{eqnarray}
\begin{eqnarray}
W^{Red} \big[ A, \Phi \big]  = \frac {1}{2}
\int d^4x \left[\left(\partial_0 A^I_i\right)^2- \left(B_i\right)^2 \right] +
\int dt
\left(I_E + I_\Phi \right).
\end{eqnarray}

Emphasize that in accordance
with the property (3.4) the field \( A^I \) at the spacial infinity can be
only stationary
\begin{eqnarray}
 A^I_i (\vec x,t )\Bigl\vert_{\mid {\vec  x} \mid~~~ \to~~~ \infty }
= \;b_i(\vec x)\;,                               .
\end{eqnarray}
that means the diagonalization of the kinetic term in the action (3.8):
(\(I_E\:=\:0\)).

The background field \( b(\vec x)\) can be treated as the zero mode of the
time derivative operator in the Gauss law \( \;\quad \partial_0 b_i^I\;=\;0 \).
As a  consequence, we have  the factorizable form for the zero mode field :
\be
\Phi(t,\vec x)^a\,\Bigl\vert_{\mid {\vec  x} \mid~~~ \to~~~ \infty }\;=
\;\varphi_0(t)\Phi_0^a (\vec x).
\ee
To clear up the meaning of \(\varphi_0(t) \)
recall that by definition the variables \(A^I\)  describe
local excitation, while zero mode field \(\Phi\) is associated with
some global (collective) dynamics of gauge fields.
Such  global properties of the theory are connected
with the well - known topological invariant \cite{BPST}, \cite{Fad}
\be
\nu\;=\;\frac{e^2}{64\pi^2}\int d^4x  F^{\nu\mu}{\tilde F}_{\mu\nu}\,.
\ee
On the constraint shell the quantity \(\nu \) can be represented as
\be
\nu^{Red}\;=\;\nu\;\Bigl\vert_{\mbox{constraint}}\;=\;
\;\int d^4x
\left( \partial_0 A_i^I +  \nabla_i \Phi\right)\;
{\bar B}_i ,\quad {\bar B}_i^a = \frac{e^2}{8\pi^2}\, B^a_i.
\ee
We can find the connection between the time - dependent part of the zero mode
\(\varphi_0(t) \) and this topological invariant after
using the following decomposition for  \(\nu\)
on the local and global parts:
\be
\nu^{Red}\;=\;\int dt \partial_0 N_T[A^I, \Phi^I]\,, \quad   \quad
N_T[A^I, \Phi ] \;=\;N_L[A^I] + N_0 \,,
\ee
\bea
N_L[A]& =& \frac{e^2}{16\pi^2}\;\; \int d^3 x \epsilon_{ijk}
\;(A^a_i\; \partial_j\; A^a_k + \frac{1}{3}\; \epsilon^{a b c}
\, A^a_i\, A^b_j\, A^c_k)\,, \\
\partial_0 N_0 &\;=&\;
\int d^3 x \bar B_i \nabla_i\,\Phi
\;=\;
\oint ds_i \left(\Phi \bar B_i\right)
\Bigl\vert_{\mid {\vec  x} \mid~~~ \to~~~ \infty } \,.
\eea
Comparing (3.14) and (3.19) we get the desirable  connection
\be
\varphi_o \;=\;\partial_0 N_0I_B^{-1},
\ee
where the constant \(I_B\) is determined via the stationary
part of fields (3.13) at spatial infinity
\be
I_B\; =\;
\oint ds_i \left(\Phi_0 \bar B_i\right)
\Bigl\vert_{\mid {\vec  x} \mid~~~ \to~~~ \infty }.
\ee
Finally, our reduced action gets the form:
\begin{eqnarray}
W^{Red} \big[ A^I, N_0 \big]&  = & W^{Red}_L[A^I] + W^{Red}_{G}\,,\nn\\
W^{Red}_L[A^I] &=& \frac {1}{2}\int d^4x \left[\left(\partial_0 A^I_i\right)^2
- \left(B_i \right)^2 \right]\,,\nn\\
W^{Red}_{G}&=& \int dt {\left(\partial_0 N_0 \right)}^2 I ,\quad
I\;=\;I_\Phi / I_B^2 . \label{eq:redaction}
\end{eqnarray}

 From the reduced action (\ref{eq:redaction}) with zero mode we get an
unexpected result:

\noindent{\it There is a static solution of pure Yang - Mills theory with
finite energy in four - dimensional Minkowskian space - time} ~\cite{Khved}.

One can show that one of  such  static solutions coincides with
 the well - known Bogomoln'yi -- Prasad --
Sommerfield (BPS) monopole solution of the Yang - Mills - Higgs system~
\cite{BPS}.
To prove this statement, it is enough to observe that the zero mode in
the action
(\ref{eq:red1}) plays the role of the Higgs stationary field.
Indeed, the
Prasad --  Sommerfield solution of the Bogomoln'yi equation
\begin{equation}
\nabla^{ac}_i (A){\Phi_0}^c = \frac{2\pi}{\mu}B^a_i (A),
\end{equation}
\begin{eqnarray}
{A}^a_i &=& \frac{1}{e} \epsilon^{abi} m^e \left[\frac{\mu}{\sinh (\mu r)} -
\frac{1}{r} \right];
\;\;\; m^l = \frac{x^l}{r}; \;\; r = \mid \vec x \mid \,,\nonumber \\
\Phi^a_0& =& \frac{2\pi}{e} m^a \left [ \coth{(\mu r)}- \frac{1}{\mu r} \right]
\,,
\end{eqnarray}
with $\mu $ being the parameter of the mass dimension automatically
satisfies the zero mode equation (3.5) and our boundary conditions.
For these field configurations, the constants (3.11), (3.21) are the
following
\begin{equation}
{I_B}\;=\;1\,, \quad \: I = I_\Phi =\frac{2 (2\pi)^3 }{\mu e^2}\,.
\end{equation}

It is interesting to note that  there are arguments~\cite{Leut}, ~\cite{Hasen}
in
favour of stability of this perturbation theory under small
deformations around this  vacuum background.

\subsection{Zero mode and  homotopy group}

It is usually assumed that all nontrivial topological properties of
the gauge theory
are connected with the existence of classical solutions of the Euclidean
Euler -- Lagrange equations with finite action --- instantons.
Recall that instanton calculations are based on consideration of the
topological nontrivial gauge symmetry group
{}~\cite{BPST}, \cite{Fad}:
\[
\hat A_\mu\,\,\to\,\, \hat A_\mu^g = g (\hat A_\mu +
\partial_\mu)\,\,g^{-1}\;\,,
\]
with restriction on  the class of stationary transformations
\(\{ g(\vec x) \}\)
with the asymptotical property
\[
\lim_{\mid {\vec  x} \mid~ \to~ \infty } g(\vec x)
= 1\,.
\]
One can use the assumption of compactification of three - dimensional
space into a three--sphere \( S^3\).
In this case all maps \(\{ g(x) \} \;\;: S^3 \;\;\to SU(2) \)
can be   split into the
disjoint homotopy classes characterized by the integer index \(n\):
\[
n = \frac{1}{24\pi^2}\;\;\int d^3 x\; \epsilon^{ijk}
tr \left[ \hat V _i\; \hat V _j\; \hat V _k \right];\;\;
\hat V_i = g\; \partial\,_i g^{-1} .
\]
Thus,  we  can speak about the homotopy group \(\pi_3 (SU(2)) = Z \).
The configurations  belonging to the  different classes cannot be deformed
continuously into each other.
The gauge transformations which are deformable to identity are called small,
while
homotopically nontrivial ones  \(n \ne 0 \) are large.
For large gauge transformations the local
topological variable $N_L[A]$ (3.18)  varies as
\be
N_L[A^g]\;=\;N_L[A]\;+\;n .
\ee
The group of these transformations is usualy considered in the context
of instanton approximation \cite{BPST} for the Green functional in the
Euclidean  space
\[
  G^{Euc}(A_1, A_2 ;T) =  \int_{A_1}^{A_2}[DA]e^{-W_{Euc}} ,
\]
according to which the contribution from the self - dual fields \((E=\pm B) \)
dominates in the vaccum sector.
One can write the spectral decomposition of this
functional
\[
  G^{Euc}(A_1, A_2 ;T) =  \sum_{\epsilon}e^{-\epsilon T}
\Psi _{\epsilon}(A_1)\Psi^{*} _{\epsilon}(A_2),
\]
with wave functions satisfying
the following set of equations
\begin{eqnarray}
H_L\Psi_{\epsilon}& = &\epsilon \Psi _{\epsilon}\,,\\
\nabla_i E_i\;{\Psi}_\epsilon & = &0\,, \\
T_L\Psi_{\epsilon}& = &e^{i\theta}\,\,\Psi_{\epsilon} .
\end{eqnarray}
The first equation is the stationary Schr\"odinger equation with the
Hamiltonian
\[
H_L[A,E] = \int d^3 x \frac{1}{2}\,(E^{a2}_i + B^{a2}_i) \,.
\]
Eq.(3.28) reflects the invariance of the theory under the small gauge
transformations, while  eq.(3.29) describes the covariance properties of the
wave
function under a large gauge transformation with the topological shift
operator $T_L$ represented in the following form :
\[
T_L = \exp\left\{ {\frac{ d }{{dN_L[A]}}} \right\} ,
\]
where $N_L $ is the functional (3.18). This form is justified in refs.
\cite{BPST}
by representing the solution of (3.27) -- (3.39) in the form
of the Bloch wave function
\[
\Psi_{\epsilon} (N_L,\,A) = e^{{i} P\;N_L}\;
\Psi_{\epsilon} ( A )
\]
with the exact  solution with energy
$\epsilon\;=\;0 $
\[
\Psi_0 = \exp\left\{ \pm \frac{8\pi^2}{e^2}\,N_L [A] \right\},
\]
which represents the quantum version of the instanton solution
\((\hat E \Psi_0 =\pm \hat B \Psi_0 )\).
However, this solution is nonphysical (nonnormalizable).
It is easy to check also that the operators $H_L, T_L $
do not commute
\[
[H_L,T_L] \ne 0,\;\quad \left[ [H_L [H_L,\;T] \right] \ne 0,\,
\]
therefore they cannot have a common system of physical eigenstates.
So, for such local realization of the topological shift operator \(T_L\)
the following statement is valid:

\noindent {No - go theorem: {\it There are no physical solutions
for  equations \/} (3.27) -- (3.29)}.

These obstacles  can be overcome in our gaugeless consideration
where the winding number
is represented according to (3.17) as a sum of the local functional $N_L$
and some collective
variable $N_0$. Introduction of this new variable  allows a consistent
description of  the representation of the homotopy group.
Indeed, starting with  reduced action one can get the following
Hamiltonian:
\begin{equation}
H^{Red} \;=\; \frac{1}{2I} {\hat P_0}^2 + H_L[A^I,E^I] ,
\end{equation}
as the sum of the local Hamiltonian $H_L$  and the global one with
canonical momentum conjugated to the global variable \(N_0\)
\be
P_0\;\equiv\; \frac{\delta W^{Red}(N_0)}{\delta\partial_0 N_0}\;=
\;\partial_0N_0.
\ee
The requirement of the invariance under the large gauge transformations
(3.26)
leads to the region of definition of the variable \(N_0 \) --- \([0,1)\).
In this case eqs. (3.27)-(3.29) transform as follows
\bea
H^{Red} \; \Psi_{\epsilon}& =& \epsilon \Psi_{\epsilon}, \nn \\
\nabla_i E_i\;{\Psi}_\epsilon & = &0\,, \\
T_G\Psi_{\epsilon}& = &e^{i\theta}\,\,\Psi_{\epsilon},
\eea
where
\begin{eqnarray}
T_G \; =\;\exp \left( i \hat P \right)\; =
\; \exp\; (\frac{d}{dN_0}).
\end{eqnarray}
These equations admit the factorization of the wave function on the plane
wave describing the topological collective motion, with the momentum
spectrum:
\[
P_0 = 2\pi k + \theta\;
\]
and on the oscillator - like part depending on transverse variables :
\begin{eqnarray}
\Psi_{\epsilon} (N_0,\,A^I) = < P_0 | N_0 > \,\Psi_{L} [A^I] \;\;\quad
< P_0 | N_0 > \;=\;e^{i P_0 N_0}.
\end{eqnarray}

Thus, the consistent solution of the problem of the quantization of the
Yang  -- Mills theory with nontrivial homotopy group is achieved by
introducing an additional global variable arising in the process of
reduction  as zero mode of Gauss' constraint.

\subsection{Zero mode and topological confinement}

Generally speaking, for constrained system
in the ``gauge"
\be
 {\nabla}^{ab}_i(A)\;\partial_0 A_i^{b}\;=\;0   \label{eq:gauge}
\ee
the generating functional for Green's function  can be constructed
{}~\cite{PervushRiv} by
using the conventional  Faddeev --
Popov functional integral ~\cite{FadSlav}:
\[
Z[ J ]\;=\;\int \prod_\mu D^4A_\mu
[\det {({\nabla}^{2}(A_i))}]
\delta ({\nabla}_i(A)\;\partial_0 A_i) e^{i W[ A ] + i
S[ J]}.
\]
After the integration over \(A_0\) it can be  rewritten as
\begin{eqnarray}
Z[ J ]\;=\;\int D^3A_i
[\det {({\nabla}^{2}(A_i))}]^{1/2}
\delta ({\nabla}_i(A)\;\partial_0 A_i)
e^{i W^{Red}[ A ] + iS[J]}.
\end{eqnarray}
The calculation of this functional integral faces the problems of
the  existence of zeroes  of FP determinant well known as the
Gribov ambiguity of fixing variables ~\cite{Gribov}.
In the gaugeless approach this ambiguity
corresponds to the existence of the  zero mode sector (\ref{eq:zero})
and, accordingly, of two types of variables (\ref{eq:small}) and
\bea
A_{k}^{\Phi}& =& v^{\Phi} ( A_{k}^I + \partial_{k} ) ( v^{\Phi})^{-1}\,,\nn\\
v^{\Phi}& =& T \exp \bigl\{ \int_{}^{t} dt'{{\hat \Phi}^I}\bigl\}\,,
\end{eqnarray}
satisfying one and the same gauge constraint (\ref{eq:gauge}).
The above - introduced variables \(A^I\) (\ref{eq:small}) are invariant
under the
small gauge transformations, while the variables \(A^\Phi\) are invariant
against the large one.
By the construction, the gauge factor \( v^{\Phi} \) neutralizes the large
gauge transformation,  as  the factor \( v^{I} \)   neutralized the small one
in (\ref{eq:small}) and the total topological variable has the form
\be
N_T\;=\;N_L[{A^I}] + N_{0} = N_L[A^{\Phi}] + {\mbox{Invariant term}}.
\label{eq:topol}
\ee
In particular, for the BPS  fields (3.24), the second  Bogoliubov
transformation  is
\be
v^{\Phi} =  e^{i N_0 {\tau}^a m^a\beta(r)}\;,
\ee
with function  \(
\beta(r) = {2\pi} [ \coth{(\mu r)}- \frac{1}{\mu r} ],
\)
and the invariant term in eq.(\ref{eq:topol}) has the following form:
\begin{equation}
\mbox{Invariant term} = - \frac{\sin (2\pi N_0)}{2\pi} .
\end{equation}

It is important to note that in the  conventional Hamiltonian approach
Gauss' constraint is
considered as the generator of small gauge transformations .
In the gaugeless approach one can be convinced that Gauss' constraint is
responsible for both
gauge transformations: small and large.
The small gauge transformations are  generated by  Gauss'
constraint without zero mode, while the large by the zero mode (3.34).

In the gaugeless  approach,
the requirement  of  gauge invariance of the  observable quantities
under the whole group of
transformations (small and large)
means that we should work in terms of variables \(A^\Phi\).
The gaugeless reduced configuration space  besides the variables
\(A^\Phi\) contains also the above - introduced topological variable
\(N_0 \) and, therefore, in the expression of the corresponding generating
functional there is an additional functional integral over it.
To write down the generating functional we must separate
the stationary asymptotic part \(b\),
which is accompanied by \(N_0 \),
>from the dynamical one
\[
\hat A^{\Phi }_k (\vec x,t) = v^\Phi\left( \hat b^I_k(\vec x) + \partial_k
+ \hat a^I_k ( \vec x, t)\right) v^{\Phi} =
v^\Phi\left( \hat b^I_k(\vec x) + \partial_k  \right)v^\Phi +
a^\Phi(\vec x,t).
\]
where
\[
(a^\Phi)^c = \Omega^{c d}({\vec x} \vert N_0(t))(a^I)^d \,,
\]
the matrix \( \Omega^{c d}({\vec x} \vert N_0(t)) \)
realizes  the transformation  (3.38)
in the adjoint representation of the color group
\[
v^{\Phi} \hat \tau^a ( v^{\Phi})^{-1}\;=\;\Omega^{a b} \tau^b.
\]
In the following we shall call \( b \) the condensate and
\( a \)  the quasiparticle excitation.

Thus, instead  of (3.37) in the gaugeless reduce scheme we have the
representation for the generating functional of Green's functions:
\begin{eqnarray}
Z_{P Q}^{Red}[ J^\Phi ]\;=\;
\int_{P_0(0)=P}^{P_0(T) = Q} DP_0 DN_0 e^{W^{Red}_G[P_0, N_0]}
\int \left[ D^3a_i\right]
e^{i W^{Red}_L[b^I+a] + i S^\Phi} , \nn\\
      \int \left[D^3a_i\right] =
\int D^3a_i[\det {({\nabla}^{2}(b^I+ a))}]^{1/2}
\delta ({\nabla}_i(b^I+a)\partial_0 a_i)
\end{eqnarray}
with
\[
S^\Phi[ J^\Phi, \Phi]\;=\;\int d^4x J_b^\Phi a_b^\Phi \, =
\int d^4x J_b^\Phi a_c \Omega^{c b}(x \vert N_0(t))\;.
\]

The generating functional (3.42) is free from zeroes of FP determinant and
corresponds to gauge invariant  Green's functions of quasiparticles
\be
G_{\Phi}(1,...,n)= < vac~ P | T\left[a^\Phi(1) \cdots a^\Phi(n)\right]| vac~
Q>.
\label{eq:Green}
\ee
In Eq.(\ref{eq:Green})  the vacuum vector \( | vac~ Q> \)  means the state
without
quasiparticles and  with definite topological momentum \(Q\).

The perturbation theory with respect to the quasiparticles in the
background \(b \) is constructed by  the decomposition of action
\be
W^{Red}= W_G + W_0 [b] +\frac{1}{2}\int d^4x
[(\partial_0 a)^2 - a (\tilde \Delta) a] + W_{int}[a,b],
\ee
where \(\tilde \Delta \) is the differential operator :
\[
(\tilde \Delta)^{cd}_{ij} = \delta_{ij}(\nabla^2(b))^{cd} +
2e F_{ij}^a(b)\epsilon^{cad}.
\]
After the introducing the complete set of eigenfuctions
\[
(\tilde \Delta)^{cd}_{ij}f_j^d(\vec x | w) =
w^2 f_i^c(\vec x | w)
\epsilon^{cfd},
\]
we can write the following  expansion for the field \( a^I \) :
\[
a^I(\vec x, t ) = \sum_{w}\left( c^+(w)f_j^d(\vec x | w)
e^{+i{w}t}
+c^{-}(w)f_{j}^d(\vec x | w) e^{-i{w}t}
\right).
\]
In the canonical operator quantization
the coefficient \( c^{+}(w) ( c^{-}(w) ) \) is the creation (annihilation)
operator of quasiparticles with the asymptotical Hamiltonian
\[
H_L^{Asympt}  = \sum_w w c_w c^+_w .
\]

The function \( f_{j}^d(\vec x|w) \) is the amplitude of  probability
to find the quasiparticle with the energy \( w \) at the point
\( \vec x \).
For observable quasiparticle \(a^\Phi \) this amplitude has the form :
\[
< vac~ P | a^{\Phi c}_j(\vec x, 0)| vac~ Q > =
\int_0^1 dN e^{i N(P-Q)}\Omega^{c d}({\vec x} \vert N_0(0)=N)
f_{j}^d(\vec x|w) \,.
\]

The  factor \( \Omega \) reflects  the  degeneration  of the
quasiparticle energy under the topological variable and leads to the sum of
Kronecker symbols \(\delta_{P,Q\pm \pi \beta(r) }\).
This result can be treated  as  confinement of color states
{}~\cite{PervushRiv}, \cite{Han}.
For colorless states topological degeneration factors disappear
and we get the conventional expressions for corresponding matrix elements.
This scheme of topological confinement will be discussed in more detail in
the forthcoming  publication .
It is worth to note that there are the values of the coupling constant
\( e^2  = 2 (\mid 2\pi  k + \theta \mid )^{-1}\) for which the background
part \(W_0[b]\) of the action (3.44)  is compensated by the collective motion
one
\( W_G \).

\newpage

\section{Summary}

We have discussed the
method of quantization of the Yang - Mills theory in which the nondynamical
variables are eliminated  by  the explicit resolving of the
classical Gauss constraint
\[
\nabla^2\, (A) A_0 - \nabla_i\, (A)
\partial_0 A_i = 0,
\]
and then quantum theory is built up  for the  action on the constraint shell.
We  take into account the zero modes of all operators in the constraint:  \\
\--- the condensate $ b(x)$ as a zero mode of the operator $\partial_0$
\[ \partial_0 b (x) = 0, \]
\--- the phase angle $\Phi (x)$ as a zero mode of the operator $\nabla^2 (A) $
\[\nabla^2(A) \Phi = 0. \]
This zero mode is associated with the collective coordinate
which restores the gauge invariance of the theory,
broken in any particular solution of this
constraint.
At this point the situation is quite similar  to the case of semiclassical
soliton quantization, based on the  introduction of collective coordinates
{}~\cite{Gervais} for taking into account the breaking of rigid global
symmetries (e.g., translation, rotation, etc.).
The collective variable  allows us to separate zero modes of the
Faddeev - Popov determinant and to solve the problem of gaugeless version
of Gribov's copies.

There is  a  significant   difference   between the
Yang - Mills theory with the collective variables and
the conventional  one.
In particular, it  has been shown that in the former there is a static stable
solution that corresponds to the well - known
Prasad - Sommerfield solution in the conventional Yang -Mills theory
interacting
with  Higgs' field. In the gaugeless reduced Yang - Mills the zero mode
\( \Phi(x) \) plays the  role of
the Higgs field .

The collective variable associated with the phase angle \(\Phi \) describes
the dynamical realization of the homotopy group \(\pi_3 (SU(2))= Z  \)
in the Minkowskian space.
In  contrast to the instanton version  with integer
winding number functional, in the dynamical realization the collective
variable as  winding number is  continuous. Just the averaging
over the continuous winding number leads to  confinement of gauge invariant
color
fields configuration due to the phenomenon of the complete
destructive interference.

{\bf Acknowledgments }

\vspace{0.3cm}

The authors thank Profs. P.I. Fomin, B. Ovrut, S. Sawada, E. Seiler,
M. K. Volkov and
Drs.  S. Gogilidze, A. Kvinikhidze, G. Lavrelashvili for useful discussions.
The work was
supported in part by the Russian Foundation  of Fundamental Investigations,
Grant No 94\--02\--14411.


\vspace{1.0cm}

\begin{thebibliography}{99}
\bibitem{Dirac59} P.A.M. Dirac,  Phys. Rev. 114 (1959) 924;
\bibitem{Slav} J.C. Taylor, Nucl. Phys. B33 (1971) 436;\\
	A.A.Slavnov, Teor. Mat. Fiz.  10 (1972) 99;
\bibitem{FadSlav} L.D.Faddeev, A.A.Slavnov, {\it Introduction in the Quantum
Theory of Gauge
Fields }, Moscow, Nauka, 1984;
\bibitem{Tav}V.V.Vlasov, V.A. Matveev, A.N. Tavkhelidze, et. al.,
Phys. of Elem. Part. Nucl. 18 (1987) 5;
\bibitem{Gribov} V.N.Gribov, Nucl. Phys. B 139 (1978) 1;
\bibitem{Singer} I.M.Singer, Commun. Math. Phys. 60 (1978) 7;\\
Physica Scripta. 24 (1981) 817;
\bibitem{Pervush1} V.N.Pervushin, Teor. Mat. Fiz.  45 (1980) 327;
\bibitem{PervushRiv} V.N.Pervushin, Riv. Nuovo Cimento 8 (1985) N 10, 1;
\bibitem{Ilieva}N.P.Ilieva, V.N.Pervushin, Sov. J. Par. Nucl. 22 (1991) 573;
\bibitem{FadJack} L.D. Faddeev, R. Jackiw, Phys. Rev. Let. 60 (1988) 1692;\\
R.Jackiw, {\it (Constrained) Quantization without tears \/}, preprint
CTP - 2215 MIT, Cambridge, 1993;
\bibitem{Han} Nguyen Suan Han, V.N. Pervushin, Fortsch. Phys. 37 (1989) 611;
\bibitem{Khved} A.M. Khvedelidze, V.N. Pervushin, {\it Zero modes of first
class
constraints in gauge theories\/}, preprint JINR E2-93-439, Dubna, 1993 ;
\bibitem{Bogol} N.N.Bogoliubov, Sov. J. Phys. JETP 1  (1958) 51;
\bibitem{BPST} A.A.Belavin et al., Phys.Lett 59 (1975) 85;\\
R.Jackiw, C.Rebbi, Phys. Lett. 63B (1976) 172;\\
R.Callan, R.Dashen, D.Gross, Phys. Rev. D17 (1977) 2717;
\bibitem{Fad}L.D.Faddeev, Proc. Fourth Intern. Symposium on
Nonlocal Field Theory, 1976, D1-9788, Dubna, (1977), 267;
\bibitem{HP} W. Heisenberg, W. Pauli, Z.Phys. 56 (1929)1; ibid
59 (1930) 166;
\bibitem{Zumino} B.Zumino, J. Math. Phys. 1 (1960) 1;
\bibitem{BPS} M.K.Prasad, C.M.Sommerfield, Phys.Rev.Lett. 35 (1975) 760,\\
  E.B. Bogomol'nyi Yad. Fiz 24 (1976) 449;
\bibitem{Leut} H. Leutwyler, Nucl.Phys B179, (1981), 129;
\bibitem{Hasen} R.Jackiw, C.Rebbi, Phys. Rev. D13 (1975) 3398;\\
A.S.Goldhaber, Phys. Rev. Lett. 36 (1976) 1122;\\
P.Hasenfratz, G.'t Hooft, Phys. Rev. Lett. 36 (1976) 1119;
\bibitem{Gervais} J.L.Gervais, A.Jevicki, B.Sakita, Phys. Rep. 23C (1976) 281;
\end{thebibliography}
\end{document}